\documentstyle[12pt]{article}
\begin{document}

\def\lsim{\:\raisebox{-0.5ex}{$\stackrel{\textstyle<}{\sim}$}\:}
\def\gsim{\:\raisebox{-0.5ex}{$\stackrel{\textstyle>}{\sim}$}\:}

\begin{flushright}
TIFR/TH/94-06 \\
hep-lat/9403001 \\
\end{flushright}
\bigskip
\bigskip
\bigskip

\begin{center}
\large {\bf FINITE TEMPERATURE PHASE TRANSITION IN SU(2) LATTICE GAUGE
THEORY WITH EXTENDED ACTION} \\

\bigskip
\bigskip
\bigskip

\large {Rajiv V. Gavai, Michael Grady\footnote {On leave from Dept.
of Physics, SUNY Fredonia, Fredonia N.Y. 14063, U.S.A.}
and  Manu Mathur\footnote {On leave from S. N. Bose National Centre for
Basic Sciences, DB-17, Salt Lake, Calcutta-700 064, India}} \\
\bigskip
Theoretical Physics Group \\
Tata Institute of Fundamental Research \\
Homi Bhabha Road, Bombay 400 005, India \\

\bigskip
\bigskip
\bigskip
{\bf ABSTRACT}\\ \end{center}
\bigskip
\noindent  We study the three dimensional
fundamental-adjoint $SU(2)$ lattice gauge theory at finite temperature
by Monte Carlo simulations.  We find that the finite temperature deconfinement
phase transition line joins the first order bulk phase transition line at its
endpoint.  Moreover, across the bulk transition line, the Polyakov
loop undergoes a discontinuous jump implying the existence of both
confining and deconfining phases on its two sides. Implications for
universality and the nature of the confining-deconfining transition
are discussed.

\newpage

\begin{center}
1. \bf INTRODUCTION \\
\end{center}
\bigskip

Quantum field theories are defined with an ultraviolet cut-off.  The
choice of this regulator is highly non-unique.  In perturbation theory the
renormalizability criterion ensures that the physical answers are
independent of the cut-off parameter used.  In fact, different choices of
the regulators only change the scale in the theory, leaving physics
unchanged.

A non-perturbative way to regulate a theory is to define it on a lattice,
which has a built-in ultraviolet cut-off (the lattice spacing $a$).
A regulated theory on the lattice corresponding to a particular
theory in the continuum is non-unique due to the large freedom of choice
of lattice action and the lattice type.
This non-uniqueness is expected to be a lattice
artifact and should disappear as the lattice spacing $a$ tends to zero
(or the lattice correlation length $\rightarrow \infty$), resuling in a
universal continuum limit.  The Wilson formulation \cite{Wil} of lattice
gauge theory is a particular choice of lattice action from a much more
general class, given by
\begin{eqnarray}
S = \sum_P \sum_R \beta_R~Re~Tr_R U_P ~~~~~~~~~~~~~~~.
\label{AA}
\end{eqnarray}
Here $U_P$, the plaquette action, is an ordered product of the four
group elements $U_{ij} ~(\in SU(N))$ corresponding to the  four links of an
elementary plaquette.  The sum over $R$ extends over all the
representations of the gauge group and the sum over $P$ runs over
all plaquettes. $Tr_R$ denotes the trace over colour space in the
representation $R$.

A variety of different forms of lattice actions have been employed
in the past to check the scaling behaviour and the universality
of the $SU(2)$ gauge theories.  Dimensionless ratios of physical
quantities, such as $M_G/\sqrt{\sigma}$\cite{MuSch}, or $T_c/\sqrt
{\sigma}$\cite{Gav}, where $M_G$ is the lightest glueball mass,
$T_c$ is the deconfinement temperature and $\sigma$ is the string
tension, have been computed for the $SU(2)$ lattice gauge theory
and shown to be almost independent of the lattice actions used.
Even more stringently, functions of such dimensionless ratios,
e.g., the energy density as a function of $T/T_c$, have been
successfully used to demonstrate\cite{GaKaSa} universality.  While these
results have boosted our confidence in the existence of a non-perturbative
continuum limit of the $SU(2)$ gauge theory, it is still not clear
whether more qualitative aspects of the theory, such as the order
of the deconfinement phase transition, also exhibit this universality.
We investigate the confining-deconfining behaviour in the extended coupling
plane of a particular class of theories belonging to eq. (\ref{AA}) and
compare it with that of the usual Wilson action.
The organization of this paper is as follows: In sect. 2  we introduce
the model and the motivation to study it at non-zero temperatures.  In
sect. 3  we give a very brief review of the observables computed and
their critical behaviour.  In sect. 4 we discuss the data and its
analysis.  Sect. 5 consists of the observations and their
implications.
\bigskip

\begin{center}
2.  \bf THE MODEL \\
\end{center}
\bigskip

A well studied model belonging to the above class of $SU(2)$ gauge theory
is described by the action
\begin{eqnarray}
S = \sum_P \left( \beta \left(1 - {1\over2} Tr_F U_P \right) +
         \beta_A \left(1 - {1\over3} Tr_A U_P \right) \right)~~~~~.
\label{AA1}
\end{eqnarray}
Here $F$ and $A$ denote the fundamental and adjoint representations
respectively.
Comparing the naive classical continuum limit of eq. (\ref{AA1})
with the standard $SU(2)$ Yang-Mills action, one obtains
\begin{eqnarray}
{1 \over g^2_u} = {\beta \over 4} + {\beta_A \over 3}~~.~~
\label{gu}
\end{eqnarray}
Here $g_u$ is the bare coupling constant of the continuum
theory.  One can also characterize the model with a pair of couplings
$(g_u,\theta)$ with $\tan\theta = \beta_A/\beta$.  In the
non-perturbative continuum limit (i.e. $a \rightarrow 0$)
each of these theories,
characterized by a given $\theta$, flow to the same critical fixed point,
$g^c_u = 0$.  The corresponding asymptotic scaling relation \cite {Kor} is
\begin{eqnarray}
a = {1 \over \Lambda (\theta)} \exp\left[- {1 \over 2\beta_0 g^2_u}\right]
\left[\beta_0 g^2_u\right]^{-\beta_1 \over 2\beta^2_0} ~~~,~~
\label{ga}
\end{eqnarray}
where
\begin{eqnarray}
\log {\Lambda (0) \over \Lambda (\theta)} = {5 \pi^2\over 11}
{ 6 \tan \theta \over \left( 3 + 8 \tan \theta\right) }~~.~~
\label{lambda}
\end{eqnarray}
Here $\beta_0$ and $\beta_1$ are the usual first two coefficients of the
$\beta$ function for the $SU(2)$ gauge theory.

This model has been extensively studied after Bhanot and Creutz \cite {Bha}
found it to have a rich phase structure (fig.1).  At $\beta_A = 0$,
it reduces to the standard Wilson action.  Along the $\beta = 0$ axis
it describes the $SO(3)$ model which has a first order phase transition
at $\beta_A^c \sim 2.5$.  At $\beta_A = \infty$ it describes the $Z_2$
lattice gauge theory again with a first order phase transition at
$\beta^c = {1\over2} \ell n(1 + \sqrt{2})$ $\approx$ 0.44 \cite {Weg}.
Bhanot and Creutz found that these first order transitions extend into the
($\beta$,$\beta_A$) plane, ending at an apparent
critical point located at (1.5,0.9).  These transition lines are shown in
fig.1 by continuous lines. The same phase diagram was also later
produced qualitatively by mean field theory  techniques \cite {Alb}.
The $\Lambda$-ratio was found to be smaller than its asymptotic value
given by eq. (\ref{lambda}) by a factor of $\sim$3.5 in a Monte Carlo
determination\cite{BhDa} from the ratio of string tension evaluated on
$\beta_A=$constant lines around (1.5, 0.9) and (2.2, 0.0)
in the ($\beta$, $\beta_A$)-plane.  It was later shown\cite{GaKaSa}
that the $\theta$ dependance of this factor further differs rather
strongly from eq. (\ref{lambda}) on the $\beta_A=0.9$ line.  Thus
in the vicinity of the endpoint of the
first order line, the asymptotic scaling relation is known to be
violated strongly. Ref. \cite{BhDa} also found that the string tension
does not go through zero in this region, suggesting the endpoint to
be a higher order phase transition, although the location of endpoint
was obtained in Ref. \cite{Bha} by extrapolating the discontinuity
in average action to zero.

The motivation for Bhanot and Creutz to study
this model by Monte Carlo simulation was to show that non-abelian lattice
gauge theories need not be free of singularities to maintain confinement
in the asymptotic scaling regime.  As fig. 1 shows, one can go smoothly
around the line of first order singularities without affecting confinement
in the asymptotic regime.  However, this is perhaps an incomplete picture
of the phase diagram of the model.  Since the numerical results, which
established the phase diagram were obtained on small, $N_\sigma^4$, lattices
for $N_\sigma = 5$-7, one must worry about a possible deconfinement
phase transition on these lattices at sufficiently small coupling $g_u$.
As all finite lattices are necessarily at finite temperature as well,
the confining-deconfining transition line must exist in the extended plane
on all lattices and one needs to understand its behaviour on larger
lattices to draw conclusions about the zero temperature phase diagram
of the theory and its possible implications to the continuum limit of the
confining phase.  The deconfinement phase transition has recently
been extensively studied at $\beta_A = 0$ \cite {Eng,Fing}.
We follow this transition in the $\beta_A$ direction
and find that it merges with the bulk phase transition line found by
Bhanot and Creutz.  The phase diagram therefore splits into seperated
confining and deconfining phases.
\bigskip

\begin{center}
3.  THE OBSERVABLES AND THEIR CRITICAL BEHAVIOUR \\
\end{center}
\bigskip

On a finite lattice of size $N^3_\sigma \times N_\tau$ the volume and
temperature are given by \cite {Sve1}
\begin{eqnarray}
V = (N_\sigma a)^3 ~{\rm and}~ T = {1 \over N_\tau a}~~~~.
\label{vt}
\end{eqnarray}
Here $a$ is the lattice spacing governed by the bare coupling $g_u$
for sufficiently small $a$, as shown in eqns.\,(\ref{ga}-\ref{lambda}).
To avoid big finite spatial volume effects, one chooses
$N_\sigma > N_\tau$ for finite temperature simulations. One basic observable
 which we studied is the average plaquette energy $\langle P \rangle$, where
\begin{eqnarray}
P = {1\over2}{{\sum_P Tr_F U_P}\over{6N_\sigma^3N_{\tau}}}
\label{plaq}
\end{eqnarray}
Due to periodic boundary conditions in the timelike direction, the $SU(2)$
lattice gauge theories described by eq. (\ref{AA}) and in particular
eq. (\ref{AA1})  have a $Z_2$ invariance at finite temperature corresponding
to the center of the group.  Under this
\begin{eqnarray}
U_0 (\vec n,\tau_0) \rightarrow z U_0 (\vec n,\tau_0) ~~\forall n,~~ \tau_0 :
{\rm fixed}~ {\rm ,~~and}~~ z ~~\in Z_2 ~~~~.
\label{sym}
\end{eqnarray}
Here $U_0 (\vec n,\tau)$ is the timelike link at the lattice
site $(\vec n,\tau)$.  This transformation
leaves the action  invariant but the Polyakov loop, defined by,
\begin{eqnarray}
L(\vec n) = {1\over2} Tr \prod^{N_\tau}_{\tau=1} U_0 (\vec n,\tau),
\label{pol}
\end{eqnarray}
changes:
\begin{eqnarray}
L \rightarrow z L~~~~ .
\label{tr}
\end{eqnarray}
A nonvanishing value for $\langle L \rangle$ signals a spontaneous
break-down of the global $Z_2$ symmetry.
The thermal expectation value of the Polyakov loop
or its average value $L = {1 \over N_\sigma^3} \displaystyle \sum_{\vec n}
L(\vec n)$ can also be shown to be the order parameter for the
deconfinement phase transition since it is a measure of the free energy
of an isolated free quark \cite{Sve1,McSv}.

It is been argued that the effective theory for this order parameter,
obtained by integrating out all other degrees of freedom, is
in the same universality class as the Ising model in 3-dimensions \cite
{Sve1,Sve2}.  Thus, if the deconfinement phase transition is of second
order, then it will have the same critical exponents as those of
the Ising model, provided the universality class does not have another
second order phase transtion.  On an infinite lattice, these exponents
are $\beta$, $\gamma$ and $\nu$ corresponding to the order parameter
itself, its corresponding susceptibility and the correlation length:

\begin{eqnarray}
\langle L\rangle & \propto & |T - T_c|^\beta  {\rm ~~~~~for}
                         ~~~~~{T \rightarrow T^+_c }\\[2mm]
\chi & \propto & |T - T_c|^{-\gamma}
               {\rm ~~~for}~~~~~{T \rightarrow T_c }\\[2mm]
\xi & \propto &  |T - T_c|^{-\nu} {\rm ~~~for}~~~~~{T \rightarrow T_c}~~~~ .
\label{critexp}
\end{eqnarray}
For the 3-d  Ising model $\beta \approx 0.325$, $\gamma \approx 1.24$
and $\nu \approx 0.63$.  For the finite temperature $SU(2)$ gauge theory,
one obtains these exponents from Monte Carlo simulations by
simply fitting the order parameter\cite{Eng,McSv,GaSa} or by using the
finite size scaling theory\cite{Barb} for the susceptibilty.
According to the latter, the susceptibility on a lattice
of spatial extent $N_\sigma$ is expected to grow like
\begin{equation}
\chi \propto N_\sigma^{\omega}~~,~~
\label{chifs}
\end{equation}
where $\omega=\gamma/\nu=1.97$ according to the universality prediction
above.  If the phase transition were to be of first order instead, then
one expects the exponent $\omega = 3$, corresponding to the dimensionality
of the space \cite {ChLaBi}.  In addition, of course, the order parameter is
expected to exhibit a sharp, or even discontinuous, jump and the corresponding
probability distribution should show a double peak structure.
For $\beta_A = 0$, the universality prediction was verified
by Monte Carlo simulation
by Engels et. al. \cite {Eng}, who found $\omega= 1.93 \pm 0.03$.

\bigskip

\begin{center}
4.   DATA AND ANALAYSIS \\
\end{center}
\bigskip

Our Monte Carlo simulations were done on $8^3 \times 4$ and $10^3 \times 4$
lattices, using the Metropolis algorithm.  Around $\beta_A = 1.1$ where the
nature of phase transition changes from 2nd order to 1st order, we also
simulated the model on a $12^3 \times 4$ lattice.  The
Ferrenberg-Swendsen \cite {Fer} technique was used to extrapolate the data
at neighbouring couplings to locate the critical point.  Once located,
much longer runs were performed at these points.  (200,000, 150,000 and
50,000 sweeps on $8^3 \times 4$, $10^3 \times 4$ and $12^3\times4$ lattices
respectively). Table 1 shows the values of the couplings ($\beta$,$\beta_A$)
where simulations were performed along with the critical couplings
($\beta_c$) determined from the peak in susceptibility using Ferrenberg-
Swendsen method.   The last entry in the table (1) is the value of
finite size scaling exponent ($\omega$) calculated from the values of the
susceptibility peak on $8^3\times 4$ and $10^3 \times 4$. At $\beta_A$=0.9
and 1.1 the second value of $\omega$ is calculated from the susceptibility
peaks on $8^3\times 4 $ and $12^3\times 4 $ lattices.


As the expectation value of the Polyakov loop is always zero on a finite
lattice, we measure the absolute value of $L$ after taking its lattice
average.  The values of average Polyakov loop
$\langle |L|\rangle$, extrapolated by the Ferrenberg-Swendsen technique
in the neighborhood of critical points are plotted in fig. 2-a,b,c,d for
all $\beta_A$ along with the actual values determined from the individual runs.
The error bars shown were determined from binning.   As seen
from fig. 2, the slope of the Polyakov loop curve keeps increasing with
increasing value of $\beta_A$.   All the corresponding  values of
susceptibilities,
\begin{eqnarray}
\chi = N^3_\sigma (\langle L^2\rangle - \langle |L|\rangle^2),
\label{ki}
\end{eqnarray}
are plotted in fig. 3-a,b,c,d.
We have also shown the first order and second order predictions
for the susceptibility peaks on the $10^3 \times 4$ and $12^3 \times 4$
lattices based on the measured susceptibility peak on the $8^3 \times 4$
lattice and $\omega=3.0$ and 1.97 respectively.
As is clear from the susceptibility  plots, the universality hypothesis is
well supported by the data for all the $\beta_A$ including and up to
$\beta_A=0.9$.  Instead of using the critical exponents above, one can try
to obtain them from the data by simply comparing the peak heights.
The corresponding exponents from
the data for susceptibility  are given in table 1.  They exhibit an
excellent agreement with the universality prediction of $\omega=1.97$
for $\beta_A \le 0.9$, confirming the second order nature of the phase
transition.

At   $\beta_A = 1.1$ there is evidence that the transition has
become first order. The histograms of $|L|$ (fig. 4) on $N_\sigma=10,12$
show a definite two peak structure with the dips between the two peaks
increasing with $N_\sigma$ . The value of the susceptibility peak,
however, lies
below the first order predicted values but definitely outside the second order
predicted region also. This is unlike the clear second order nature of the
transition up to $\beta_A = 0.9$.  It is possible that
at $\beta_A = 1.1$, the corrections to the leading scaling prediction
of eq.\,(\ref{chifs}) are large and thus larger lattices are needed to
ascertain the true nature of the phase transition.  It has been
found in simulations of Potts models with first order transitions, that much
larger lattices are needed to see the correct finite size scaling exponents
than are needed to see a two peak histogram \cite {Bil}. To verify the
onset of the first
order nature of the phase transition for $\beta_A \ge 1.1$,  we also simulated
the model at $\beta_A = 1.5$, where we find it to be strongly first order.
We performed $10^5$ sweeps at $\beta = 1.05$, and 40,000 sweeps
at $\beta = 1.02$ and 1.0535 (fig. 5-a,b,c).  Due to the very large
tunnelling time ($>10^5$ sweeps), we were unable to locate the exact critical
coupling and verify the first order scaling relation. But the evolution curves
for both $|L|$ and $P$ at $\beta = 1.02$, $1.05$ and $1.0535$
clearly indicate the
co-existence of two phases seperated by a large barrier. The intermediate runs
at $\beta=1.03$
(50,000 sweeps) and at $\beta=1.04$ (100,000 sweeps) also did not show any
tunnelling.  The data at $\beta_A = 1.1$ and $1.5$ also indicates that the
Polyakov loop has a discontinuity exactly across the bulk transition line,
coincident with the discontinuity in the average value of the plaquette.
The discontinuity in the values of average Polyakov loop (plaquette energy)
at $\beta_A = 1.1$ and 1.5 is 0.23 (0.04) and 0.48 (0.25) respectively.

\begin{center}
5.  DISCUSSION  \\
\end{center}
\bigskip

 From fig. 1 and from the results presented in the previous section, it is
clear that  for $\beta_A \simeq 1.0$, the deconfinement phase transition
joins the bulk transition line, and surprisingly remains
 apparently coincident for all values of $\beta_A \geq 1.0$.
As far as, we can tell from our simulations, there is really no evidence
for two transitions.  The discontinuity in both the average plaquette
and the average Polyakov loop is located at the same $\beta$ and the
latter jumps from an essentially zero value to a rather large value at
the same point.  The fluctuations in both the phases are rather small
and are not suggestive of any superimposed second order phase transition,
although the dominant first order nature
could make it difficult to see them.
The scaling behaviour of the finite temperature and bulk transitions
is different for $N_\tau, N_\sigma \rightarrow \infty$. The
$\beta_c$ for the former should
move to infinity, whereas it should remain
anchored at finite value in the latter case.
Therefore the joining together of these two
transitions is a curious phenomenon, which, if it persists in this limit,
would lead to a paradox. In the following, we discuss three possible
scenarios and their implications as one approaches the continuum limit of
the theory.

\begin{enumerate}

\item[{A]}] The most conservative possibility is that the joining of the
two lines is accidental for $N_\tau=4$ but they will eventually separate out
for large enough $N_\tau$.  This scenario is consistent with Bhanot and
Creutz's interpretation and the observed universality of the deconfinement
phase transition from the equality of its scaling exponents with those
of the 3-d Ising model.  To explore this possibility a
little further, we simulated the model at $\beta_A=1.1$ and 1.5 on a
$12^3\times 6$ lattice also.
Due to the large simulation time we could only bracket the  critical
coupling by looking at the peak position of the susceptibility and also
the behaviour of the histograms.  Our estimated $\beta_c$ for $\beta_A$=1.1,
$N_\tau=6$ is
$1.3425 \pm 0.0025$ which should be compared with the critical coupling on
$8^3\times4$ lattice, $\beta_c =1.3270 \pm 0.0008$, obtained from our data
on the susceptibilty peaks\footnote{The errorbars given are upper limits based
 on the couplings of nearby runs which were seen to be definitely confined and
 deconfined.}. Therefore, the shift in  $\beta_c$ in
going from $8^3\times 4$ to $12^3 \times 6$ lattice
at $\beta_A= 1.1$  is definitely less than 0.0188.
Moreover the transitions still remained coincident and first order. At
$\beta_A =1.5$, due to the large tunnelling time even on the $8^3\times4$
lattice, we ran a hysteresis cycle instead of runs at individual couplings.
The forward hysteresis cycle at $\beta_A = 1.5$, on $12^3 \times 6$ shows
$\beta_c < 1.065$. On the other hand,  on the  $8^3 \times 4$ lattice
the evolution curves (fig. 5-a,b,c) show that the $\beta_c > 1.02$.  Therefore
the shift in $\beta_c$ on going from $N_\tau =4$ to $N_\tau = 6$ is also
less than 0.045 here. Recall that at $\beta_A =0$ this shift was observed
to be 0.13\cite {Fing}, and if the asymptotic scaling relation were to be
valid in this region, the expected shift in the critical coupling is 0.15.
Of course, the scaling relations, eqs. (\ref{gu}-\ref{ga}) are known to be
strongly violated in this region.  Nevertheless, the shifts are too tiny and at
least suggest that much larger lattices are required to confirm this scenario
if one is to move away from the bulk transition and see a separate
second order deconfinement phase transition.

\item[{B]}] Taking the coincidence of finite temperature deconfinement phase
transition and the bulk phase transition more seriously, one has basically two
extreme alternatives.  Either the interpretation of Bhanot and Creutz\cite{Bha}
of the bulk transition line was incorrect, and one has only a deconfinement
transition to deal with, or the identification of the deconfinement transition
at $\beta_A=0$ needs to be investigated more,
and one has only a bulk transition
line on which the order of the transition changes from first to second at
the point (1.5, 0.9).  Both the alternatives have their own problems and
inconsistencies with the published results.  Thus, the nonzero string
tension reported in Ref. \cite{BhDa} around this point suggests a higher order
phase transition at that point while the results on the growth of
susceptibility, reported both in this work, and Ref.  \cite{Eng} are indicative
of a second order phase transition for $\beta_A \le 0.9$.
Indeed, if the entire phase transition line is
bulk then all the evidence for the universality class of the deconfinement
phase transition at $\beta_A=0$ will have to be treated as accidental and
a priori the global $Z_2$ center symmetry has also no unique role to play
in that case.
This is because for a four dimensional bulk system there is no particular
reason to choose periodic boundary conditions, from which the $Z_2$ symmetry
arises, whereas this is required for the finite temperature interpretation.
However it should be pointed out that the use of an asymmetric lattice will
mask the true scaling behavior of a 4-d bulk system, because the system will
behave as a thin film or layer system which will have a three dimensional
critical behavior in the limit $N_\sigma \rightarrow \infty$, $N_\tau$
finite \cite{Fisher}. From this point of view it is not surprising that
the system exhibits a 3-d critical behavior, typified by the 3-d Ising model,
in this limit, even if the underlying transition is a 4-d bulk transition
which would presumably have a completely different critical behavior
in the limit $N_\tau = N_\sigma \rightarrow \infty$.
Of course, a crucial prediction of this hypothesis is that
the entire transition line will persist at finite $\beta_c$ as
$N_\tau \rightarrow \infty$ for {\em all} $\beta_A$, and no connection
of the confining phase of the pure gluonic theory with its
asymptotically free phase will be possible.  Therefore
the continuum limit will
be always deconfining,
a possibility which has been raised previously in several contexts\cite{nocon}.

On the other hand, if the entire line is a
deconfinement phase transition line, it will move towards the right
hand side of the phase diagram as $N_\tau \rightarrow \infty$.  However,
our observation of an extremely small shift in $\beta_c$ as $N_\tau$ went
from 4 to 6 at $\beta_A = 1.1$ makes it unlikely to observe this shift
for any $N_\tau$ used in any numerical simulations
up to now.  Moreover, the large discontinuity in the plaquette
and the Polyakov loop then suggests the deconfinement phase transition to
be of first order at $\beta_A=$1.1 and 1.5.   Thus, the universality
class seems to change with an apparently irrelevant coupling on these
lattices.  One can then only hope that the universality will be restored
on larger lattices.

\end{enumerate}

\bigskip

To summarize, we have studied the finite temperature deconfinement phase
transition for the extended $SU(2)$ action on $N_\sigma^3 \times
N_\tau$ lattices with $N_\sigma = 8$, 10, 12 and $N_\tau= 4$ and 6.  We
found that as $\beta_A$ increases, the deconfinement phase transition
moves towards smaller $\beta$ and appears to join the previously known
bulk transition line.  The universality of the critical exponents for the
$SU(2)$ deconfinement phase transition thus seems to be violated
since the finite temperature deconfinement phase transition appears to
change from being a second order one to a first order transition unless
this behaviour is purely accidental on small lattices.
Future simulations on larger lattices could, of course, yield a separation
of the bulk transition from the deconfinement phase transition, however,
no such hints could be obtained in the limited variations studied here.
It will be interesting to study whether a similar phenomenon
also takes place for the $SU(3)$ theory.  We conclude that much larger
lattices are required to settle the issues raised in this paper.

\newpage

\pagestyle{empty}

\begin{table}
\begin{center}
{Table 1}
\end{center}
The values of ($\beta,\beta_A$) at which simulations were performed,
$\beta_c$ and the \\
finite size scaling exponent $\omega$.  The expected value for
$\omega$ is 1.97 (3.0) if \\
the deconfining phase transion is second order (first order). \\ \\
\medskip
\begin{tabular}{|c|c|c|c|}
\hline
                         &                   &                       &       \\
{}~~~~~~$\beta_A~~~~~~~~$ & $~~~~~~~\beta~~~~~~~$ & $~~~~~~~~\beta_c~~~~~
{}~~~$ & $~~~~~~~~~\omega~~~~~~~~~$ \\
                         &                   &                       &       \\
\hline \hline
    &      &                         &     \\
    & 1.830 ~~~$:~8^3\times4$ &1.831~  &     \\
0.5 &      &                         &     \\
    & 1.830 ~~~$:~10^3\times4$ &1.830~   &  1.92(29)\\
    &      &                        &          \\
\hline
    &      &                        &          \\
     &1.610 ~~~$:~8^3\times4$  &1.613~  &      \\
0.75 &       &                      &       \\
     &1.610 ~~~$:~10^3\times4$   &1.609~ &  1.53(32) \\
    &      &                        &          \\
\hline
    &      &                        &          \\
     &1.489 ~~~$:~8^3\times4$ &1.486~ &        \\
     &       &                       &         \\
0.9  &1.489 ~~~:$~10^3\times4$ &1.485~  &  2.24(64)\\
     &       &                        &          \\
     &1.489 ~~~:$~12^3\times4$ &1.486~  &  2.10(22) \\
    &      &                        &          \\
\hline
    &      &                        &          \\
    &1.32620~~~:$~8^3\times4$ &1.3276~    &         \\
    &                       &                         &         \\
    &1.32635~~~:$~8^3\times4$ &1.3275~    &         \\
    &                       &                         &         \\
    &1.32700~~~:$~8^3\times4$ &1.3270~    &         \\
1.1 &                       &                         &         \\
    &1.32650~~~:$~10^3\times4$ &1.3273~   &         \\
    &                       &                         &          \\
    &1.32700~~~:$~10^3\times4$ &1.3271~   & 2.31(28) \\
    &                       &                         &          \\
    &1.32685~~~:$~12^3\times4$ &1.3270~  & 2.34(15) \\
    &      &                        &          \\
\hline
\end{tabular}
\label{tabgas}
\end{table}

\newpage
\pagestyle{empty}

\begin{center}
\bf{FIGURE CAPTIONS}
\end{center}

\bigskip

\noindent Fig.1
The phase diagram of the extended SU(2) lattice gauge theory.
The solid lines are from simulations done on a $5^4$ lattice by
Bhanot and Creutz\cite{Bha}.  The broken
line is the finite temperature deconfinement phase transition line
based on the results discussed in the paper.
\bigskip

\noindent Fig.2
Average Polyakov loop ($\langle |L| \rangle$) at (a) $\beta_A$=0.5, (b)
$\beta_A$=0.75,
(c) $\beta_A$=0.9 and (d) $\beta_A$=1.1  on $8^3 \times 4$ and $10^3 \times 4$
lattices.  At $\beta_A=0.9$ and $1.1$, results on $12^3 \times 4$
lattice are also shown. The points with error bars are results of simulations
and the curves are extrapolations by the Ferrenberg-Swendsen technique from
the run closest to the peak.
\bigskip

\noindent Fig.3
The susceptibility curves at (a)$\beta_A$=0.5, (b)$\beta_A$=0.75, (c)$\beta_A$
=0.9 and (d)$\beta_A$= 1.1 on $8^3 \times 4$, $10^3 \times 4$ and $12^3 \times
4$ lattices.  The first order (F.O) and second order (S.O) predictions,
explained
in the text,  are explicitly shown. The higher (lower) predictions are for the
$12^3\times4$ ( $10^3\times4$ ) lattice. The points
and curves are as in Fig. 2.
\bigskip

\noindent Fig.4
The probability density of $|L|$
at $\beta_A =1.1$ on $8^3\times 4$ ($\beta=1.32635$), $10^3\times4$ ($\beta=
1.3265$) and $12^3 \times 4$ ($\beta=1.32685$) lattices.
\bigskip

\noindent Fig.5
Evolution of $|L|$ and (P+0.15) on $8^3 \times 4$ lattice,
at $\beta_A=1.5$ and (a)$\beta =1.02$, (b)$\beta=1.05$, (c)$\beta=1.0535$.
The upper broken lines are cold starts and the lower solid lines are
hot starts.

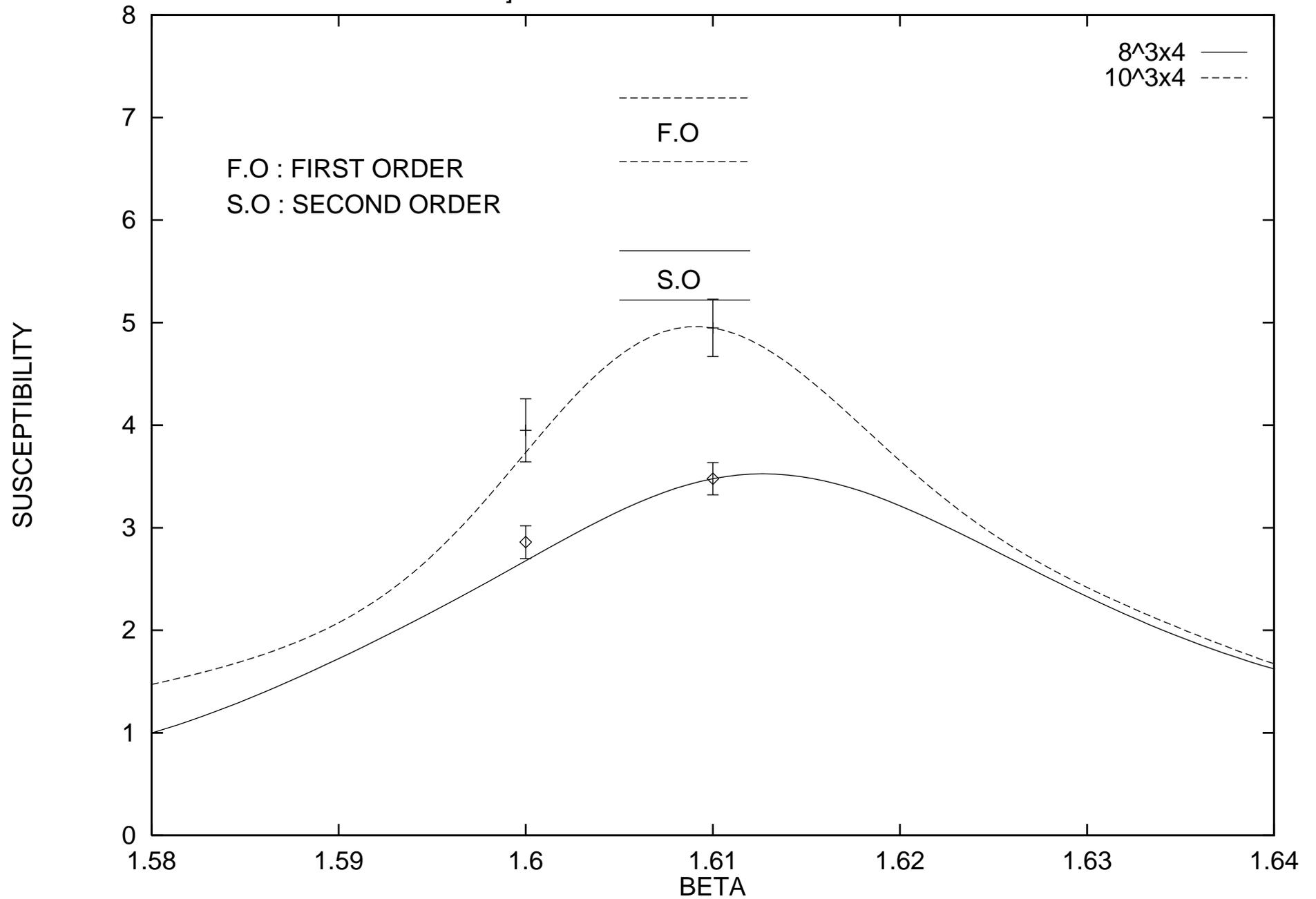
\end{document}